\documentstyle[epsfig,psfig]{texas}

\begin{document}

\title{Numerical hydrodynamics on light cones}
\author{Jos\'e A. FONT and Philippos PAPADOPOULOS}
\address{Max-Planck-Institut f\"ur Gravitationsphysik,
Albert-Einstein-Institut, \\
Schlaatzweg 1, Potsdam, D-14473, Germany
\\
{\rm Email: font@aei-potsdam.mpg.de, philip@aei-potsdam.mpg.de}}

\begin{abstract}

Characteristic methods show excellent promise in the evolution of 
single black hole spacetimes. The effective coupling with matter 
fields may help the numerical exploration of important astrophysical 
systems such as neutron star black hole binaries. To this end we 
investigate formalisms for numerical relativistic hydrodynamics 
which can be adaptable to null (characteristic) foliations of the spacetime. 
The feasibility of the procedure is demonstrated with
one-dimensional results on the evolution of 
self-gravitating matter accreting onto a dynamical black hole.

\end{abstract}

\section{Introduction}

The so-called {\it characteristic} formulation of the field equations of 
general relativity has 
proven recently to be well suited for numerical evolutions of single black hole 
(vacuum) spacetimes~\cite{BGLMW1}. A three-dimensional characteristic code developed 
by the Binary Black Hole Grand Challenge Alliance in the US has achieved, for the 
first time, accurate and long-term stable evolutions of such spacetimes, including 
distorted, moving and spinning single black holes, with evolution times up to 
$60,000M$~\cite{60000M}. Such evolutions are ultimately possible due to the excellent 
gauge control properties furnished by null foliations, gauge control being a major
open issue in the more usual 3+1 formulation.

The incorporation of matter terms into such a computational framework, in the form of 
some suitable stress-energy tensor, permits the study of interesting astrophysical 
scenarios with optimal computational efficiency. Such an approach allows to investigate 
the non-linear evolution of self-gravitating matter around {\it dynamic} black holes. 
Some interesting astrophysical studies to be performed include, e.g., accretion disks 
or the non-linear and fully relativistic evolution of a compact binary system formed 
by a black hole and a neutron star. This binary system is one of the 
anticipated sources of gravitational radiation. Its detailed numerical study can 
provide valuable information on the waveform signals and a means to decipher the 
nuclear equation of state.

In this report we focus on the formulation of the hydrodynamic
equations on null foliations of the spacetime. The reader must be
aware that most of the previous analytic and numerical work in
general relativistic hydrodynamics (GRH hereafter) has been done using 
the ``standard" 3+1 formulation, i.e., using a spacetime foliation of spacelike
hypersurfaces (see, e.g.,~\cite{BFIMM} and references therein).
Additionally, the small amount of existing work dealing with the numerical
integration of the GRH equations on null hypersurfaces can be characterized by: a) 
a direct way of writing the hydrodynamic equations with no attention to important 
mathematical (and numerical) details (as, e.g., hyperbolicity) and b) the use of 
straightforward finite-differences numerical schemes~\cite{IWW},\cite{DDV},\cite{BGLMW2}. 
However, on the basis of the experience gained in the last few years in the integration 
of the hydrodynamic equations in the spacelike case, we believe that the procedure 
previously adopted for the null case is likely to be inadequate when dealing with 
non-trivial simulations. Most notably we refer to realistic astrophysical scenarios 
containing high speed  (ultrarelativistic) flows and steep gradients or even shock waves.

As a way to avoid these problems we propose the use of the so-called
{\it high-resolution shock-capturing} schemes (HRSC in the following)
to integrate the GRH equations on light cones.
HRSC schemes are commonly used in numerical integrations of the Newtonian
inviscid Euler equations. In recent years they have been succesfully
extended to relativistic hydrodynamics, both in the special and general 
cases~\cite{MIM},\cite{font1},\cite{EM},\cite{BFIMM},\cite{FMST}. Mathematically, 
HRSC schemes rely on the hyperbolic character of the hydrodynamic equations. The 
knowledge of the characteristic fields (eigenvalues) of the system, together 
with the corresponding eigenvectors, allows for accurate integrations, by means 
of either exact or approximate Riemann solvers, along the fluid characteristics. 

Among the interesting properties of HRSC schemes we stress the important fact that 
they are written in conservation form. Hence, all physically conserved quantities 
of a given partial differential equation will also be conserved in its finite-differenced 
version. More precisely, the time update of the following hyperbolic one-dimensional 
partial differential system of equations
\begin{eqnarray}
\frac{\partial {\bf u}}{\partial t} + 
\frac{\partial {\bf f}}{\partial x} = {\bf s}
\end{eqnarray}
is done according to the following algorithm:
\begin{eqnarray}
    {\bf u}_{i}^{n+1}={\bf u}_{i}^{n}-\frac{\Delta t}{\Delta x}
    (\widehat{{\bf f}}_{i+1/2}-\widehat{{\bf f}}_{i-1/2}) +
    \Delta t {\bf s}_i.
\end{eqnarray}
In these expressions ${\bf f}$ and ${\bf s}$ are vector-valued functions,
the fluxes and sources, and they both depend on the state vector of
the system ${\bf u}$. Index $i$ labels the grid location and $n$ the time.
Time and space discretization intervals are indicated by $\Delta t$ and
$\Delta x$, respectively. The ``hat" in the fluxes is used to denote the 
so-called numerical fluxes which, in a HRSC scheme, are computed according to
some generic flux-formula. It adopts the following functional form:
\begin{eqnarray}
  \widehat{{\bf f}}_{i\pm{1\over 2}} = \frac{1}{2}
          \left( {\bf f}({\bf u}_{i\pm{1\over 2}}^{L})  +
                 {\bf f}({\bf u}_{i\pm{1\over 2}}^{R}) -
          \sum_{\alpha = 1}^{p} \mid \widetilde{\lambda}_{\alpha}\mid
          \Delta \widetilde {\omega}_{\alpha}
          \widetilde {r}_{\alpha} \right).
\end{eqnarray}
\noindent
Notice that the numerical flux is computed at cell interfaces ($i\pm1/2$).
Indices $L$ and $R$ indicate the left and right sides of a given interface.
The sum extends to $p$, the total number of equations. Finally, quantities
$\lambda$, $\Delta\omega$ and $r$ denote the eigenvalues, the jump of the characteristic 
variables and the eigenvectors, respectively, computed at the cell interfaces 
according to some suitable average of the state vector variables. Further technical 
information can be found in, e.g.~\cite{font1} and references therein.

HRSC schemes are also known for giving stable and sharp discrete shock
profiles. They have also a high order of accuracy, typically second order
or more, in smooth parts of the solution.

\section{A covariant form of the general relativistic hydrodynamic equations}

The conservation equations in covariant form are
\begin{eqnarray}
\frac{\partial}{\partial x^{\mu}} \sqrt{-g} J^{\mu} & = & 0 \\
\frac{\partial}{\partial x^{\mu}} \sqrt{-g} T^{\mu\nu} & = & - \sqrt{-g}
\Gamma^{\nu}_{\mu\lambda} T^{\mu\lambda}
\end{eqnarray}
with the matter current and stress energy tensor for a perfect fluid given by 
$J^{\mu} = \rho u^{\mu} $, $ T^{\mu\nu} = \rho h u^{\mu} u^{\nu} + p g^{\mu\nu}$, 
where $\rho$ is the density, $p$ the pressure and $u^{\mu}$ the fluid four velocity 
obeying the normalization condition $u_{\mu} u^{\mu} = -1$. Additionally, 
$h = 1 + \epsilon + p/\rho$ is the relativistic specific enthalpy, with $\epsilon$ 
being the specific internal energy.

The construction of an initial value problem requires the introduction
of a coordinate chart $(x^{0},x^{i})$, where the scalar $x^{0}$ represents a foliation 
of the spacetime with hypersurfaces (coordinatized by $(x^{1},x^{2},x^{3})$) on 
which an appropriate initial data set is prescribed. Upon introducing the coordinate 
chart, we define the coordinate components of the four-velocity 
$u^{\mu} = (u^{0}, u^{i})$. The velocity components $u^{i}$, together with the 
rest-frame density and internal energy, $\rho$ and $\epsilon$, provide a unique 
description of the state of the fluid and are called the {\em primitive} variables.  
They consitute a vector in a
five dimensional space $ {\bf w}^{A} = (\rho, u^{i}, \epsilon)$. The index A is 
taken to run from zero to four, coinciding for the values (1,2,3) with the 
coordinate index $i$.

We define the initial value problem in terms of another vector in the
same fluid state space, namely the {\em conserved variables} ${\bf
U}^{B}$, individually denoted $(D,S^{i},E)$, as follows:
\begin{eqnarray}
D     & = & {\bf U}^{0} = J^{0}  = \rho u^{0} \\
S^{i} & = & {\bf U}^{i} = T^{0i} = \rho h u^{0} u^{i} + p g^{0i}  \\
E     & = & {\bf U}^{4} = T^{0t} = \rho h u^{0} u^{0} + p g^{00}
\end{eqnarray}
\noindent
With these definitions the equations take the standard conservation law form
\begin{equation}
\partial_{x^{0}} (\sqrt{-g} {\bf U}^{A})
+ \partial_{x^{j}} (\sqrt{-g} {\bf F}^{jA})
= {\bf S}^{A}
\label{eq:cons-law}
\end{equation}
where ${\bf S}^{A}({\bf w}^{B})$ are source terms that depend on metric
derivatives and the (undifferentiated) stress energy tensor,
$\sqrt{-g}$ is the volume element associated with the four-metric and
we define the flux vectors ${\bf F}^{jA}$
\begin{eqnarray}
{\bf F}^{j0} & = &  J^{j}  = \rho u^{j}   \\
{\bf F}^{ji} & = &  T^{ji} = \rho h u^{i} u^{j} + p g^{ij}  \\
{\bf F}^{j4} & = &  T^{j0} = \rho h u^{0} u^{j} + p g^{0j}
\end{eqnarray}
\noindent
The source terms take the form
\begin{eqnarray}
{\bf S}^{0} & = & 0 \\
{\bf S}^{i} & = & - \sqrt{-g} \, \Gamma^{i}_{\mu\lambda} T^{\mu\lambda}  \\
{\bf S}^{4} & = & - \sqrt{-g} \, \Gamma^{0}_{\mu\lambda} T^{\mu\lambda}
\end{eqnarray}

For a complete description of the local characteristic structure of the 
previous equations, which is the basic ingredient to make use of HRSC schemes, 
we refer the interested reader to~\cite{PF2}. We end this section by stressing 
that the previous formulation of the GRH equations is fully covariant and, hence, 
both applicable to spacelike or null spacetime foliations. The results
we show next are specialized to null foliations.

\section{Results}

\subsection{The Riemann problem on a null surface}

This section is devoted to demonstrate the functionality of our approach
in a standard hydrodynamical testbed computation, the Riemann problem. This is 
one of the standard tests to calibrate numerical schemes in fluid dynamics. 
It provides a strong check of our procedure in the limiting case of Minkowski 
spacetime. Initial data are given on the light cone and the evolution is compared 
to the exact solution. The initial setup consists of a zero velocity fluid having 
two different thermodynamical states on either side of an interface. In our setup 
the initial state is specified by $p_r=13.3$, $\rho_r=10$ at the right side of the 
interface and $p_l=0.66\cdot10^{-6}$, $\rho_l=1$ at the left side. We use a perfect 
fluid equation of state, $p=(\Gamma-1)\rho\epsilon$ with $\Gamma=5/3$. When the 
interface is removed, the fluid evolves in such a way that four constant states occur.  
Each state is separated by one of three elementary waves: a shock wave, a contact 
discontinuity and a rarefaction wave.  

\begin{figure}[t]
\centerline{\psfig{figure=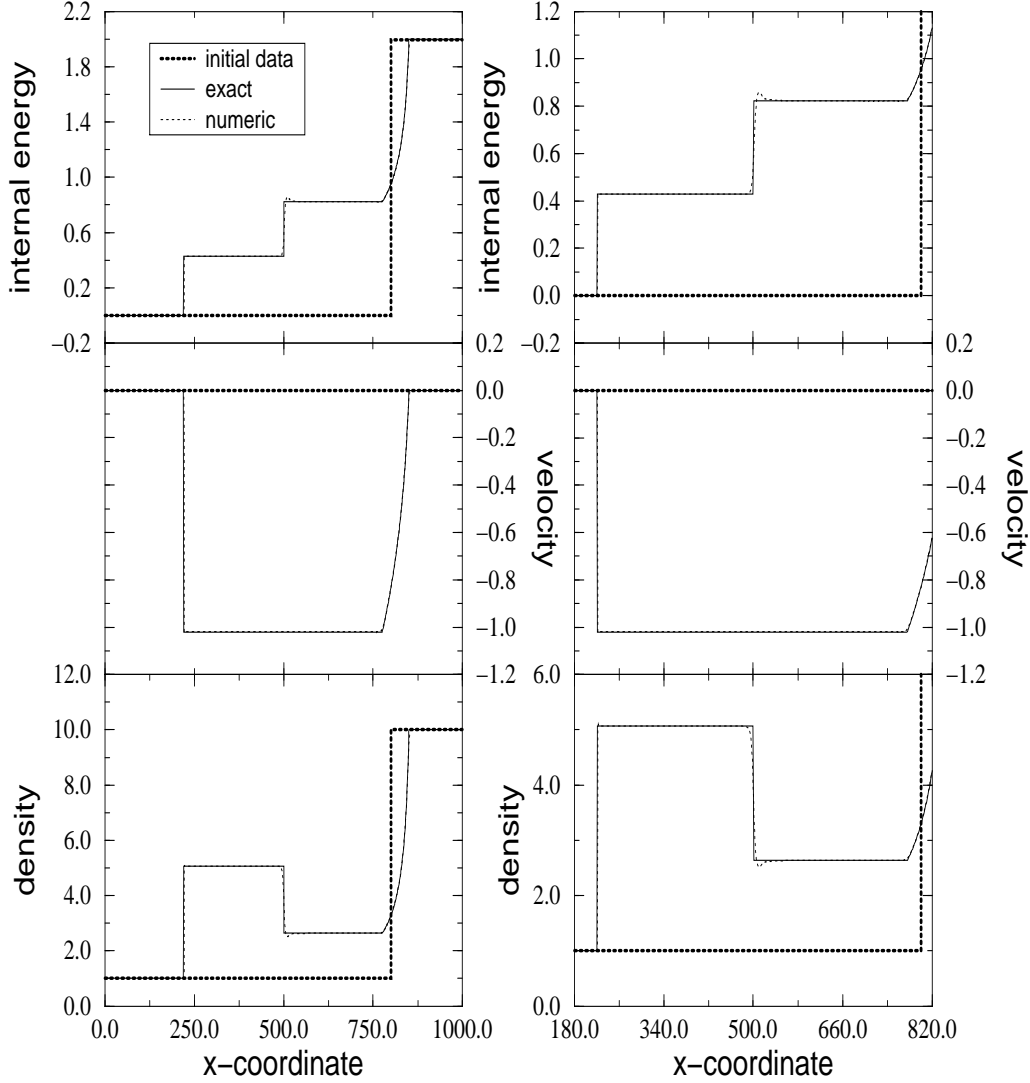,width=5.5in,height=5.8in}}
\caption{{
The shock tube problem. Exact versus numerical solution at a retarded time $v=120$
for the internal energy (top), velocity (middle) and density (bottom).
}}
\label{fig:shock}
\end{figure}

In Fig.~\ref{fig:shock} we plot the results of the simulation. The
left pannels show the whole domain, with the $x$-coordinate ranging
from 0 to 1000 whereas the right pannels show a closed-up view of the
most interesting region ($x\in[180,820]$). We use a numerical grid of 2000 
zones and, hence, a rather coarse spatial resolution ($\Delta x=0.5$).
The initial data are evolved up to a final time $v=120$. From top to bottom 
Fig.~\ref{fig:shock} displays the internal energy ($\epsilon$), the velocity ($u^x$) 
and the density ($\rho$). The thick dotted line represents the initial discontinuity. 
The solid line shows the exact solution after a retarded time $v=120$ and the 
thin dotted line indicates the numerical solution at this same time. The agreement between 
the two solutions is remarkable. The flow solution is characterized by an (ingoing) 
shock wave (moving to the
left) and an (outgoing) rarefaction wave (moving to the right). We note that those 
features of the solution moving towards the left are more developed than those moving 
towards the right.  The major numerical deviations from the exact solution appear at 
the post contact discontinuity region, with the density and
the internal energy being slightly under and over estimated there.

\subsection{Spherical accretion onto a black hole}

We further test our procedure by increasing the complexity of the 
numerical simulation with the inclusion of gravity. To this end, we consider the 
problem of spherical accretion of a perfect fluid onto a black hole.

We consider the general spherically symmetric spacetime with perfect fluid matter, 
following the formalism of Tamburino-Winicour~\cite{tamburino}.  The use of a 
spacetime foliation $v$ by null hypersurfaces ($\nabla^{\mu} v \nabla_{\mu} v = 0$), 
together with the assumption of spherical symmetry implies that the non-trivial 
Einstein equations split in the following group of equations
\begin{eqnarray}
G_{vr} & = & \kappa T_{vr}, \\
G_{rr} & = & \kappa T_{rr}, \\
G_{vv}|_{\Gamma} & = & \kappa T_{vv}|_{\Gamma},
\end{eqnarray}
where the last equation is a {\em conservation} condition to be imposed on the 
boundary $\Gamma$ of the integration domain. Adopting the Bondi-Sachs form of the 
metric element,
\begin{equation}
ds^2 = - \frac{e^{2\beta}V}{r} dv^2
+ 2 e^{2\beta} dv dr + r^2 (d\theta^2 + \sin\theta^2 d\phi^2),
\end{equation}
the geometry is completely described by the functions $\beta(v,r)$ and $V(v,r)$. 

The first two equations above, Eqs. (16) and (17), are {\em hypersurface} equations, 
to be integrated away from $\Gamma$ along the null cone. The $\beta$ hypersurface equation 
is
\begin{equation}
\beta_{,r} = 2 \pi r T_{rr}
\end{equation}
and the $V$ hypersurface equation,
\begin{equation}
V_{,r} = e^{2\beta} (1 - 4 \pi r^2 (g^{AB}T_{AB} - T)).
\end{equation}

Boundary conditions $\beta(v)_{\Gamma},V(v)_{\Gamma}$ for the radial
integrations are provided by the third equation, after adopting a
suitable gauge condition. This condition fixes the only remaining
freedom in the coordinate system, namely the rate of flow of
coordinate time at the world-tube. 
We refer the reader to~\cite{PF2} for a discussion on suitable gauge conditions.

The initial data consist of the fluid variables $(\rho,\epsilon,u^{r})$
on the initial slice $v_0$, together with the values of
$\beta(v_0)_{\Gamma}$ and $V(v_0)_{\Gamma}$.

In the limit of a test fluid, i.e., a perfect fluid with sufficiently low density 
accreting spherically onto a {\it static} black hole, the numerical 
evolution can be compared to an exact solution. This comparison was performed 
and the agreement was found to be excellent. We refer the interested 
reader to~\cite{PF2} (see also~\cite{PF1}) for further information. We focus here 
on the case of spherical accretion onto a dynamic black hole.

\begin{figure}
\centerline{\psfig{figure=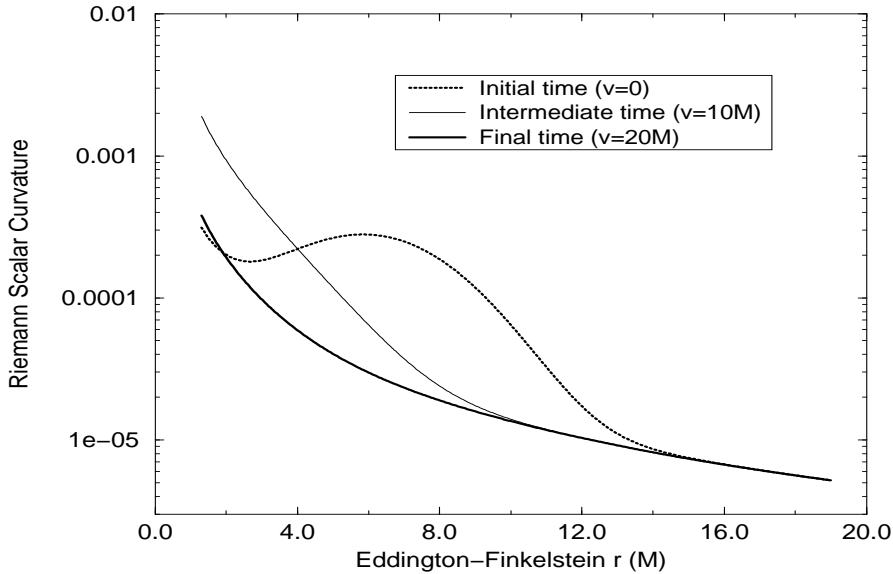,width=5.0in,height=3.5in}}
\caption{{
Spherical accretion of a self-gravitating perfect fluid: evolution of the Riemann 
scalar curvature. Once the shell is swallowed by the central object the solution is 
dominated by the curvature of the final black hole.
}}
\label{fig:accretion2}
\end{figure}

\begin{figure}
\centerline{\psfig{figure=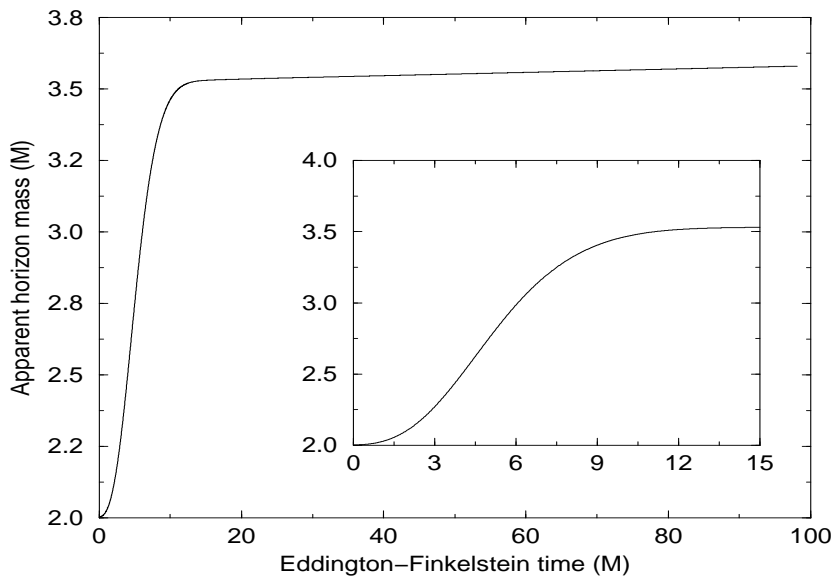,width=5.0in,height=3.5in}}
\caption{{
Spherical accretion of a self-gravitating perfect fluid: evolution of the black hole 
apparent horizon mass.
}}
\label{fig:accretion3}
\end{figure}

As initial data we consider the exact, stationary solution corresponding 
to the test fluid case and, on top of this background solution, we build a 
self-gravitating spherically 
symmetric shell whose density is parametrized by a Gaussian profile with suitable
width and amplitude.
The results of the simulation are plotted in Figs.~\ref{fig:accretion2}
and~\ref{fig:accretion3}. We find that
the shell is radially advected (accreted) towards the hole in the first $10M-15M$. 
Once the bulk of the accretion process ends, we are left with a quasi-stationary 
background solution. This is clear in
Fig.~\ref{fig:accretion2}, where we plot the evolution of the 
logarithm of the Riemann scalar curvature, as a function of the ingoing 
Eddington-Finkelstein radial coordinate, $r$.
This variable encodes a large 
amount of relevant metric information. We note that the initial shell has a 
non-negligible curvature. After the shell is accreted by the hole the curvature 
profile monotonically increases towards the central singularity. The accretion 
of the shell originates a rapid increase on the mass of the apparent horizon of 
the black hole. This is depicted in Fig.~\ref{fig:accretion3}. The horizon almost 
doubles its size during the first $10M-15M$ (this is enlarged in the insert of 
Fig.~\ref{fig:accretion3}). Once the main accretion process has finished, the mass 
of the horizon slowly increases, in a quasi-steady manner, whose rate depends on the 
mass accretion rate imposed at the world-tube, $\Gamma$, of the integration domain. 
We emphasize the fact that this numerical solution can be evolved as far as desired 
into the future. It has proven to remain remarkably stable and free of any kind of 
numerical pathologies, including secular instabilities or exponentially growing modes. 

\section*{Acknowledgements}
It is a pleasure to thank Ed Seidel for his continuous encouragement during the 
course of this work and Jos\'e M. Ib\'a\~nez for helpful discussions. 
J.A.F acknowledges financial support from a TMR grant from the 
European Union (contract nr. ERBFMBICT971902).

\end{document}